\documentclass[apj]{emulateapj}

\slugcomment{The Astronomical Journal, in press}

\shorttitle{Luminosity of the GRB 050904 prompt flash}
\shortauthors{Kann et al.}

\begin{document}

\title{The prompt optical/near-infrared flare of GRB 050904: the most luminous transient ever detected}

\author{
D.~A.~Kann\altaffilmark{1},
N.~Masetti\altaffilmark{2},
S.~Klose\altaffilmark{1}
}


\altaffiltext{1}{Th\"uringer Landessternwarte Tautenburg,  Sternwarte 5,
D--07778 Tautenburg, Germany}

\altaffiltext{2}{INAF, Istituto di Astrofisica Spaziale e Fisica Cosmica di Bologna,
via Gobetti 101, I-40129 Bologna, Italy}


\begin{abstract}
With a redshift of $z=6.295$, GRB 050904 is the most distant gamma-ray burst ever
discovered. It was an energetic event at all wavelengths and the afterglow was
observed in detail in the near-infrared bands. We gathered all available optical
and NIR afterglow photometry of this GRB to construct a composite NIR light curve
spanning several decades in time and flux density. Transforming the NIR light curve
into the optical, we find that the afterglow of GRB 050904 was more luminous at
early times than any other GRB afterglow in the pre-\emph{Swift} era, making it at
these wavelengths the most luminous transient ever detected. Given the intrinsic
properties of GRB 050904 and its afterglow, we discuss if this burst is markedly
different from other GRBs at lower redshifts.
\end{abstract}


\section{Introduction}

The detection of an extremely bright prompt optical flash accompanying
GRB 990123 \citep{Akerlof1999} highlighted the
possibility of using GRB afterglows as backlighting sources to probe
the high-redshift universe and the reinonization era
\citep{Lamb2000, Inoue2004, Totani2006}. This promise was
finally fulfilled with the discovery of GRB 050904 by the \emph{Swift}
satellite \citep{Cusumano2006a, Cusumano2006b}, lying at a redshift
of $z=6.295$ \citep{Kawai2006}. Not only was a bright
NIR afterglow discovered for this GRB \citep{Haislip2006, Tagliaferri2005}, a
prompt flash contemporaneous to the $\gamma$-ray emission was seen by
the robotic 25cm TAROT telescope \citep{Boer2006}.

Had the prompt optical flash of GRB 990123 been at a redshift of $z=1$
and had it been unextinguished by any dust, it would have peaked at an
apparent magnitude\footnote{In this paper, we use WMAP
concordant cosmology with $H_0=71$ km s$^{-1}$ Mpc$^{-1}$,
$\Omega_{\rm M}=0.27$, $\Omega_\Lambda=0.73$
\citep{Spergel2003}. For $z=6.295$, this leads to a distance
modulus of $44.11$ mag.} $R=7.6^{+0.04}_{-0.05}$ \citep[][henceforth K06]
{PaperIII}. This corresponds to an absolute $R$-band
magnitude $M_R=-36.5$ \citep[$2.3\cdot10^{16}L_{\odot R}$, assuming the
$V-R$ color for a G2V star from][]{Ducati2001},
making this the most luminous optical source ever detected at that
time. In this paper, we compile all available data on the afterglow of
GRB 050904 and construct a synthetic light curve for a hypothetical
perfectly ionized universe showing no Lyman dropout up to the redshift of the GRB. We then derive light curve
parameters and use the spectral energy distribution (SED) of the
afterglow to extrapolate the afterglow light curve into the $R$ band
assuming $z=1$. We find that the
prompt flash of GRB 050904 was the most luminous optical/NIR transient
ever detected, even exceeding the $R$-band luminosity of GRB 990123 at
peak by more than one magnitude.


\begin{deluxetable*}{lccccccccccc}
\tablecolumns{11}
\tabletypesize{\scriptsize}
\tablecaption{Results of the composite $YJHK$ light curve fitting}
\tablehead{
\colhead{fit\tablenotemark{a}}   &
\colhead{$\chi^2$} &
\colhead{d.o.f.}  &
\colhead{$m_c$\tablenotemark{b}}  &
\colhead{$\alpha_0$}  &
\colhead{$\alpha_1$}  &
\colhead{$\alpha_2$}  &
\colhead{$t_{b1}$\tablenotemark{c}}  &
\colhead{$t_{b2}$}  &
\colhead{$n_1$\tablenotemark{d}}  &
\colhead{$n_2$}}
\startdata
1 & 59.2 & 17 & $18.80\pm0.47$ & $1.39\pm0.13$ & $0.92\pm0.05$ & \nodata      & $0.35\pm0.14$ & \nodata      & $-10$    & \nodata         \\
2 & 59.7 & 17 & $20.65\pm0.17$ & \nodata      & $0.85\pm0.08$ & $2.45\pm0.18$ & \nodata      & $2.63\pm0.37$ & \nodata & $1.82\pm1.02$   \\
H06 & \nodata & \nodata & \nodata & $1.36^{+0.06}_{-0.07}$  & $0.82^{+0.08}_{-0.21}$ & \nodata & $\approx0.5$      & \nodata & \nodata & \nodata \\
T05 & \nodata & \nodata & \nodata & \nodata  & $0.72^{+0.15}_{-0.20}$ & $2.4\pm0.4$ & \nodata      & $2.6\pm1.0$ & \nodata & \nodata \\
\enddata
\tablenotetext{a}{The first fit uses data from 0.1
to 3 days after the burst, while the second fit uses only data after
0.3 days. H06 denotes the results from \cite{Haislip2006}, and T05 those of \cite{Tagliaferri2005}.}
\tablenotetext{b}{The apparent $J$-band magnitude at the respective break
time is denoted as $m_c$ (cf. Z06, Eq. 1).}
\tablenotetext{c}{The break at which the forward shock afterglow begins to dominate
over the prompt flash emission is $t_{b1}$, while $t_{b2}$ denotes
the jet break time. Break times are given in units of days after the
burst trigger.}
\tablenotetext{d}{$n_1$ and $n_2$ are the break smoothness parameters
of the first and second breaks, respectively.}
\label{LC}
\end{deluxetable*}

\section{The composite light curve of the GRB 050904 afterglow}
\label{Sec2}
\subsection{Data mining and fitting methods}
\label{Sec21}
The optical/NIR afterglow of GRB 050904 was observed by different
telescopes over almost five decades in time and more than six decades
in flux density. Due to its very high redshift, it was not detected at
wavelengths shorter than the $I_C$ band. We compiled all data from the
following papers: \cite{Haislip2006}, \cite{Tagliaferri2005}, \cite{Boer2006},
\cite{Kawai2006}, \cite{Gendre2006} and \cite{Berger2006}. All data
were corrected for Galactic extinction assuming $E_{(B-V)}=0.060$ \citep{SFD1998},
and then transformed into AB magnitudes \citep{ABMag}. To transform the
magnitudes into flux densities, we used the zero points given in
\cite{Haislip2006} to transform their data and standard zero
points for all other data. Data in \cite{Kawai2006},
\cite{Berger2006} and $z^\prime$ data from \cite{Tagliaferri2005}
are already AB magnitudes.

The highest data density is available in the $J$ band. Therefore, we started with the $J$ band
to construct a synthetic light curve by achromatically shifting the
other bands to the $J$-band light curve. This method is justified, as no
systematic color evolution was found in the data set. For the normal forward shock
evolution of the afterglow (once the cooling frequency $\nu_c$ has
passed the optical/NIR bands) such achromacy is expected. While the
early steep-to-shallow transition at $t=0.35$ days may be chromatic, we only have $J$
band data at this time anyway. In the following, we will always
assume an achromatic evolution of the afterglow.

We fitted the $J$-band data with a broken power-law \citep[for more details
on the fitting procedure, see][henceforth Z06]{PaperII}.
As \cite{Haislip2006} have noted, the early
$J$-band data before 0.3 days are brighter than the extrapolation of
later data to these times, indicating the steeper decline from what
could be a reverse shock flash. Initially, we excluded these early data from
our fit. We used the derived light curve parameters from the fit to the
$J$-band light curve as a reference light curve and fitted the $H$-band
data with this reference curve. We thus found a color $J-H$. Assuming
$J-H=const.$, we shifted
the $H$-band data to the $J$ zero point, merging the light curves. We
then fitted the combined $JH$ light curve again and shifted the $H$-band data
points in small increments until $\chi^2$ is minimized, thus deriving
the true $J-H$ color and the light curve parameters of the joint light curve which spans a
longer time range (0.3 to 12 days in the observer frame) and gives a better constraint on the post-jet break
(after 2.6 days) decay slope $\alpha_2$ (it is $F_\nu\propto t^{-\alpha}$).
The $K$-band data (where we make no distinction between $K$, $K_s$ and
$K^\prime$) and the $Y$-band data were added in the same way. These additional
shifts were not larger than 0.03 mag.

We transformed the HST NICMOS F160W (AB) data point from \cite{Berger2006}
to $H$ (AB) by using the conv\_AB value from the $hyperZ$
package\footnote{http://webast.ast.obs-mip.fr/hyperz/} \citep{hyperz},
$m_{AB}(H)=m_{AB}(F160W)+0.069$. As the final HST data point is corrected for host
contribution, we also corrected the other data points in the same way. From the
synthetic host spectrum presented in \cite{Berger2006}, we derived for the
host galaxy a $J$-band flux density of 0.055 $\mu$Jy, which transforms into
$J_{AB}=27.0$. We assumed conservative errors of 0.3 mag. The effect
of host subtraction is very small, at most 0.03 mag, since the last
afterglow data point at 5 days is still 4 magnitudes brighter than the
host galaxy.

\begin{figure}
\includegraphics[width=8.65cm,angle=0,clip=true]{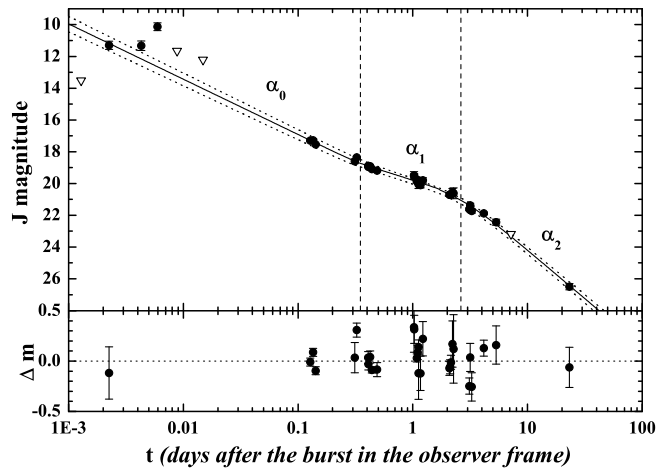}
\figcaption{Composite $z^\prime YJHK$ light curve of the afterglow of
  GRB 050904, as it has been constructed with the procedure outlined
  in \S \ref{Sec21} and \ref{Sec22}. Inverted triangles show significant upper
  limits. The solid line represents the fit with the two broken
  power-laws, the dotted lines are the $1 \sigma$ confidence interval
  of the fit. The vertical dashed lines indicate the times of the two
  light curve breaks at 0.35 and 2.63 days, and the three decay slopes
  are labeled. The residuals $\Delta m$ represent observed minus fitted
  magnitudes. The residuals have been zoomed to show the scatter in the
  data. The afterglow is seen to peak at $J\approx10$, almost three
  magnitudes above the extrapolation of the fit from 0.2 days, assuming
  a simple power-law decay.}
\label{J}
\end{figure}

\subsection{Fitting the light curve}
\label{Sec22}
The $YJHK$ composite light curve consists of an early steep
decay \citep{Haislip2006}, which goes over into a typical afterglow
decay, followed by a further break, which is identified as a jet break
\citep{Tagliaferri2005}. We fitted the two breaks separately, using
the equation from Z06 (fixing the host flux to zero as the data have
been corrected for host contribution). For the first fit, we excluded
all data beyond 3 days from the fit. The break smoothness parameter
$n$ (cf. Z06) had to be fixed to $n=-10$ (a negative value, as the
break is from steep to shallow). The jet break was fitted with $n$ left
as a variable in the fit. For this fit, data earlier than 0.3 days were
excluded. We thus derived three decay constants, which we label
$\alpha_0$ (the steep early decay), $\alpha_1$ and $\alpha_2$ (the
typical pre- and post-jet break decay). We then transformed all data
from AB back to Vega magnitudes using standard zero points.

The $z^\prime$ data \citep{Boer2006,Haislip2006,Tagliaferri2005,Kawai2006}
are already affected by Ly$\alpha$ damping. While
we used the composite reference light curve to find the $z^\prime-J$ color,
we did not implement the $z^\prime$ data in the derivation of the light
curve parameters. We also did not implement the subluminous $Z$
measurement from \cite{Haislip2006}. \cite{Boer2006} give the TAROT
measurements both as flux density at 9500 {\AA} and as $I_C$ magnitudes.
We used the flux density and transformed it to $z^\prime$ magnitudes.
Comparing these with the $I_C$ magnitudes of \cite{Boer2006}, we find
$I_C-z^\prime=2.49$ mag. This is in very good agreement with the
$I_2-z^\prime$ color of 2.48 mag derived from late afterglow data at
1.2 days, where $I_2$ is the VLT FORS2 $I$-band filter \citep[for
details on the $I_1$ and $I_2$ filters, see][]{Tagliaferri2005}.
The $z^\prime-J$ color derived at late times was then used to shift all $z^\prime$
data points to the $J$ zero point (as always, we assume achromacy,
i.e., a constant spectral slope). The result is a final composite
light curve which includes all data from the early prompt emission to
the late HST detection. In the following, when we speak of the $J$-band light
curve, we always refer to this composite light curve.

\subsection{Results of the light curve fitting}
The results of the light curve fits using the $YJHK$ light curve are given in Table \ref{LC}, and the
complete composite light curve is shown in Fig. 1
. As expected, $\alpha_1$ derived from the two different fits (cf. \S \ref{Sec22}) agrees
within errors. There is some scatter in the light curve data which leads
to the high values of $\chi^2$. This is either due to additional
measurement uncertainties or small deviations from a power-law decay,
as it has been found that the X-ray afterglow was extremely variable
\citep{Cusumano2006a, Cusumano2006b, Watson2006}. The values we derived are in full
agreement with those of other authors (Table \ref{LC}). These light curve
parameters are typical for afterglows (see Z06 for a compilation of light
curve parameters of all pre-\emph{Swift} afterglows). The light curve
steepening $\Delta\alpha=\alpha_2-\alpha_1=1.6\pm0.2$ is high but also
not extraordinary (Z06, their Fig. 3). We note that this is one of the few
light curves that allow the smoothness of the jet break to be fitted
as a free parameter, and the result is in full agreement with the
potential relationship between $\alpha_1$ and $n$ found by Z06 (their Fig. 8).

For our fit, the extrapolation to very early times meets the earliest
TAROT detection at 0.002 days, similar to what \cite{Boer2006} find.
However, the following plateau phase around 0.001 days and the flare are much brighter than
the extrapolated light curve (1 and almost 3 magnitudes,
respectively). This implies that the decay after the optical flare
must be steep. Taking into account the upper magnitude limit at 0.09 days \citep{Gendre2006}, we find
$\alpha_{flare}\geq 3$. Such a steep decay implies that the flare
could be either due to reverse shock emission or internal shocks from
late central engine activity \citep{Boer2006, Gendre2006, Wei2006, Zou2006}.
Since this is even steeper than $\alpha_0$,
an additional break must exist, but no data are available during this
time span. In particular, we find that the afterglow peaks\footnote{
The exposure time of the data point is 140 seconds \citep{Boer2006}.
Possibly, the afterglow was even brighter on shorter timescales.} at
$J\approx10$, which is among the brightest NIR afterglows ever
detected and is a first hint at its extreme luminosity.


\begin{figure}
\center
\includegraphics[width=8.65cm,angle=0,clip=true]{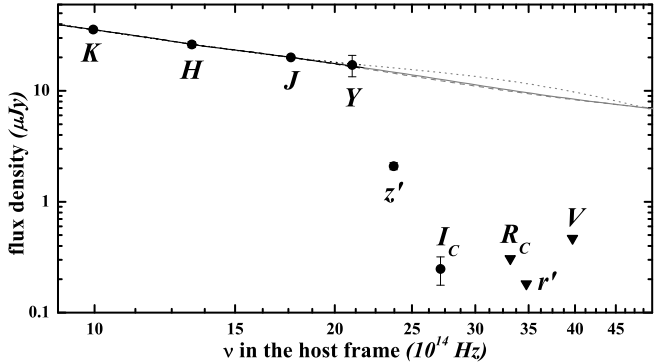}
\figcaption{The spectral energy distribution of the GRB 050904 afterglow. The three dust model fits are shown. While MW (solid) and LMC (dashed) fits are indistinguishable, the slight upturn of the SMC fit (dotted line) is seen. The Lyman absorption cutoff is not modeled, the gray fit curves are extrapolations. Only upper limits (downward pointing triangles) have been been found in the $R_C$, $r^\prime$ and $V$ bands. The flux density scale is arbitrarily chosen.}
\label{spec}
\end{figure}

\begin{deluxetable}{clcc}
\tablecolumns{4}
\tabletypesize{\scriptsize}
\tablecaption{Results of the spectral energy distribution fitting.}
\tablehead{
\colhead{source\tablenotemark{a}}   &
\colhead{dust model}   &
\colhead{$\beta$} &
\colhead{$A_V^{host}$}}
\startdata
$*$ & none & $1.00\pm0.09$ & $\;\; 0$             \\
$*$ & MW   & $0.99\pm0.10$ & $\;\; 0.02\pm0.08$ \\
$*$ & LMC  & $0.92\pm0.35$ & $\;\; 0.05\pm0.20$ \\
$*$ & SMC  & $1.31\pm1.20$ & $-0.10\pm0.40$ \\
H06 & none & $1.25^{+0.14}_{-0.15}$ & $\;\; 0$             \\
T05 & none & $1.25\pm0.25$ & $\;\; 0$             \\
T05 & none & $1.2\pm0.3$ & $\;\; 0$             \\
P06 & none & $0.3\pm0.6$ & $\;\; 0$           
\enddata
\tablenotetext{a}{$*$: This work; H06: \cite{Haislip2006}; T05: \cite{Tagliaferri2005} (two results from two different photo-$z$ fitting procedures), P06: \cite{Price2006}.}
\label{SED}
\end{deluxetable}

\section{The luminosity of the prompt flash}
\label{Sec3}

\begin{figure*}
\center
\includegraphics[width=16.5cm,angle=0,clip=true]{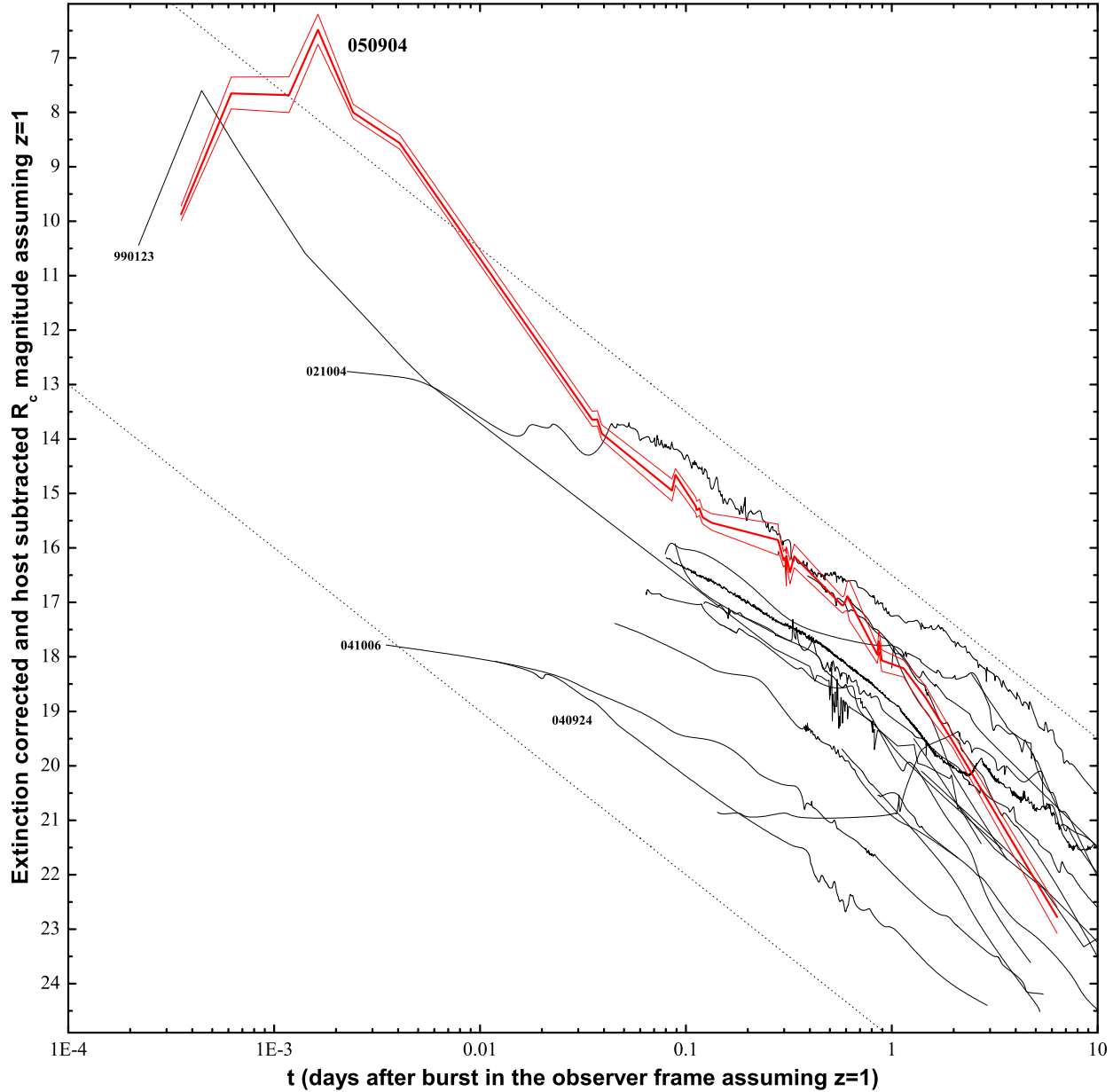}
\figcaption{Extrapolated light curve of the afterglow of GRB 050904
  (thick line) assuming a fully ionized universe with no Lyman dropout, corrected for
  extinction by dust in our Galaxy and in the GRB host galaxy and shifted to $z=1$, in
  comparison with the afterglows of the Golden Sample of K06. The
  light curve has been extrapolated into the $R_C$ band as described
  in \S \ref{Sec34}. The thin lines are the
  $1\sigma$ errors of the afterglow data, combined from measurement and
  extrapolation errors. The afterglow of GRB 050904 is more luminous than
  any other GRB afterglow at early times, making it the most luminous
  optical/NIR transient ever detected. At peak, it was about six magnitudes
  brighter than the early light curve of GRB 021004, which was the brightest
  afterglow at late times. GRB 021004 was also six magnitudes brighter than
  the extrapolation of the early afterglow of GRB 041006. At 0.05 to 0.8 days,
  the afterglow of GRB 050904 is among   the most luminous afterglows, but
  later on, it is comparable to most other afterglows. The slanted dotted lines (with decay slope $\alpha=1.2$)
  are meant to guide the eye.}
\label{z}
\end{figure*}

\subsection{Results of the spectral energy distribution fitting}

To construct the SED, we used the colors $J-H$, $J-K$ and $Y-J$ that were derived in \S 
\ref{Sec21} ($Y$, $J$, $H$ and $K$ data are not affected by Lyman $\alpha$
damping). The SED is
fitted as described in K06. First, we find an observed spectral slope
$\beta_0$. Then the SED was fitted with an additional dust extinction
curve, using dust models for the Milky Way, the Large and the Small Magellanic
Cloud \citep{Pei1992}. This gives us the extinction in the $V$ band
in the host galaxy frame $A_V^{host}$ and an intrinsic spectral slope $\beta$,
corrected for the host extinction.

To extend the SED further into the optical bands (in the observer frame), we
used reported $R_C,r^\prime,V$ upper limits. By
shifting the composite light curve to these limits, we produced additional
$R_C-J$, $r^\prime-J$ and $V-J$ colors which are lower limits on these colors.
Instead of a flat or downward curved SED, we actually found a slight upward
curvature, mainly due to the $Y$-band data (which have large errors) being
slightly too bright \citep[Fig. 2, this can also be seen in Figs. 2 and 3 of
][]{Haislip2006}. This results in a fit with SMC dust, which is
typically favored for GRB host galaxies (K06, and references therein),
finding a negative extinction, which is unphysical. The best result is
actually obtained for Milky Way dust, even though this does not imply that the
host galaxy of GRB 050904 has dust resembling that of the Milky
Way. While SMC dust is characterized by strong far-ultraviolet
extinction, MW dust has a much flatter extinction curve at these
wavelengths. Indeed, \cite{Maiolino2004} have found
dust with flat FUV extinction in a quasar at almost the same redshift
as GRB 050904, and \cite{Stratta2005} find that the
optical and X-ray afterglow of GRB 020405 can be explained by dust
with similar properties, although this GRB lies at a much closer
redshift. The characteristic 2175 {\AA} bump of MW dust is not visible
due to the very low extinction value we derive. We thus conclude that
the extinction in the host galaxy is very low and the dust is
dissimilar to SMC-type dust. In the following, we will use the values
derived from the MW dust fit, deeming this as a conservative approach.

The results of the SED fitting are given in Table \ref{SED}. Once
again, our fits are in good agreement with those of other
authors, except for \cite{Price2006}, who ascribe the low value
they derive to possible short term fluctuations. \cite{Haislip2006} and 
\cite{Tagliaferri2005} also note that while
source frame dust extinction may be present, it must be low. The
hydrogen column density along the line of sight in the host galaxy
derived via optical spectroscopy is large
\citep[log $N_H^{host}=21.3$,][]{Kawai2006}. Coupled with the very
low extinction, this implies a dust-to-gas ratio that is even lower
than for the SMC. The result is similar to GRB 990123, and in this case
also, the prompt flash may have burned dust along the line of
sight (cf. Fig. 6 in K06, and references therein).

\subsection{The circum-burst environment of GRB 050904}
Knowledge of the light curve parameters and the intrinsic spectral
slope allows us to derive conclusions on the nature of the circum-burst
environment via closure relations \citep{Price2002, Berger2002}.
We find that using $\alpha_1$ and
$\beta$, a model is strongly favored where the cooling break frequency
$\nu_c$ lies redward of the source frame ultraviolet after 0.4 days. This is
substantiated by the spectral slope of the late X-ray afterglow, which
is identical to that in the ultraviolet within errors. \cite{Watson2006}
find $\beta_X=0.88\pm0.04$, \cite{Cusumano2006a, Cusumano2006b}
find $\beta_X\approx0.9\pm0.1$. Furthermore, \cite{Frail2006} used their radio
detections of the afterglow of GRB 050904 to perform broadband
modeling and arrived at the same conclusion. In this case, one cannot draw
any conclusion on whether the GRB environment is of constant density
(ISM) or is shaped by a progenitor wind. The observed very soft jet
break ($n_2$, Table \ref{LC}) may be indicative of a wind environment
\citep{CL1999}, whereas \cite{Frail2006} use an ISM model. Finally, \cite{Gendre2006} propose that the afterglow passes through a termination shock somewhere before 0.5 days in the observer frame, switching from an early wind medium to a late ISM medium. Given the
location of $\nu_c$ redward of the optical/NIR bands, an electron spectral index of
$p=2\beta=1.97\pm0.29$ is found, which is typical for GRB afterglows
\citep[e.g.,][and references therein]{ZhangMeszaros2004}. The standard
fireball model predicts $p=\alpha_2$ \citep{ZhangMeszaros2004}, consistent
with the $\alpha_2$ we find within $1.4\sigma$ (Table \ref{LC}).

\subsection{Derivation of the $R_C$-band light curve}

In the observer frame $R_C$ band, the afterglow was unobservable due to
Lyman damping. Typically, GRB afterglows have the highest
data density in the $R_C$ band, and all afterglows that were examined
in K06 were $R_C$-band afterglows. Because of the synchrotron nature
of the afterglow radiation (described by a power-law,
$F_\nu\propto\nu^{-\beta}$, with $\beta$ being the spectral slope), and
assuming $\beta=$const., we can
construct the $R_C$-band light curve from the $J$-band light curve. The
spectral slope we derived from the observer frame NIR data can be extrapolated into
the observer frame optical, allowing us to find a $R_C-J$ color without the damping
influence of Lyman $\alpha$. For this purpose, we used the observed
spectral slope that has not yet been corrected for host galaxy extinction,
$\beta_0$. Since the host galaxy extinction is very low, the spectral curvature is
negligible and we can use a simple power-law for the
extrapolation. This power-law has an error from the fitting
process, which we added in quadrature to the measurement errors of the
data points.

\subsection{The $R_C$-band luminosity of the prompt flash}
\label{Sec34}

In order to place the afterglow of GRB 050904 in the context of the
afterglow sample discussed in K06, we shifted the afterglow to a redshift
$z=1$. Using the derived intrinsic spectral slope $\beta$, the source
frame extinction $A_V^{host}$ (both from the MW dust fit) and the redshift, we computed the magnitude
shift denoted $dR_C$ in K06. This shift takes into account the
extinction correction, the redshift difference and cosmological
$k$-corrections \citep{Sandage1988} when moving the source from its original redshift to
$z=1$. The magnitude shift creates a further error (from
$\Delta\beta$ and $\Delta A_V$) which we again added in quadrature to
the errors of the data. We find $dR_C=-5.05^{-0.13}_{+0.08}$ mag, which is higher than
any value found in K06. Applying this correction to the $R_C$-band
afterglow, we finally arrive
at our main result: had GRB 050904 lain at $z=1$ and been unaffected
by any line-of-sight extinction (including Lyman $\alpha$ damping),
the optical flare detected by TAROT would have peaked at
$R_C=6.48^{+0.27}_{-0.28}$, almost visible to the naked eye. This is
more than one magnitude brighter than the reverse shock peak of GRB
990123 (the flux density is higher by a factor of 2.8), making the
afterglow of GRB 050904 the most luminous optical transient ever
detected. \cite{Boer2006}, neglecting $k$-correction, find a peak flux density of
1080 mJy (in the $V$ band) for the afterglow of GRB 990123 and 1300 mJy (in
the $I_C$ band) for the afterglow of GRB 050904
after scaling it to the redshift of GRB 990123, but do not remark upon
this further except to call the two events comparable. For $\beta=0.99$, the absolute
$R_C$-band magnitude is $M_R=-37.6\pm0.3$ ($6.4\cdot10^{16}L_{\odot R}$),
making the afterglow in the $R_C$ band alone between 60 and 600
times brighter than the intrinsic bolometric luminosity of the
gravitationally lensed hyperluminous BAL quasar APM 08279+5255, the
most luminous persistent source known \citep{Ibata1999}.

The complete composite $R_C-$band light curve is shown in comparison with other
afterglows in Fig. 3
. It is immediately apparent that the
optical afterglow of GRB 050904 was much brighter than any other
afterglow over almost two decades in time. At the peak of the prompt
flash, it was six magnitudes brighter than GRB 021004, which is the
brightest afterglow at late times, and another six magnitudes brighter
than GRB 041006, one of the faintest GRBs in the sample. This span is
much larger than at one day after the GRB, where the afterglow
luminosities tend to cluster \citep[K06;][]{LZ2006, Nardini2006}.
At later times, though, it
resembles other afterglows. At one and four days after the burst
(Fig. 3), we find $R_C=18.13$ and $R_C=21.50$, respectively,
which transforms to absolute magnitudes $M_B=-24.51\pm0.20$ and
$M_B=-21.14\pm0.20$, respectively (cf. K06, their Figs. 9 and 10). This ranks it among the
brightest afterglows at one day, but it is not the brightest, as it has
also been noted by \cite{Tagliaferri2005}. At
four days after the burst, it lies close to the center of the
luminosity distribution.

\subsection{The spectral slope during the flare}

Since no multi-color data for the prompt emission phase are available,
we shifted the early optical emission achromatically, assuming no
spectral changes between this phase and the late afterglow phase
(where no color evolution is evident). The assumption of achromacy for
the first 100 seconds after a GRB is surely an
oversimplification. Strong spectral changes are expected within this
time \citep{SPN1998}. However, does this imply that the true
luminosity of the flare was actually lower than what we have derived?

The spectral slope of the late optical/NIR afterglow, consistent with
the cooling break lying at longer wavelengths than the optical, is
very steep, $\beta\approx1$. During the prompt emission, the forward
shock afterglow is expected to be in the fast cooling phase
\citep[and references therein]{Wu2005}. If the medium
surrounding the GRB progenitor is of constant density (ISM model), 
the optical band lies between the cooling frequency $\nu_c$ (at longer
wavelengths) and the peak frequency $\nu_m$, the expected spectral
slope is $\beta=0.5$. In case even the cooling frequency has not yet
passed the optical bands, the spectral slope in the optical actually
rises, with $\beta=-1/3$ \citep[and references therein]{ZhangMeszaros2004}.
A similar situation
would be seen if the early optical emission is dominated by the low
energy tail of the prompt emission, a situation which was seen for GRB
041219A and GRB 050820A \citep{Vestrand2005,
Vestrand2006}. This has also been suggested for GRB 050904
\citep{Wei2006}, although the optical flux in the flare is
much brighter than the extrapolation of the X-ray spectrum for GRB
050904 \citep{Boer2006}, in contrast to the cases of
GRB 041219A and GRB 050820A \citep{Vestrand2005, Vestrand2006}.
We thus conclude that the spectrum in the optical
bands during the prompt emission was not steeper than at late times,
and possibly shallower, i.e. $\beta<1$.

On the one hand, this implies that the $R_C$ magnitude of the
afterglow at $z=6.295$ would be brighter than what we derived with the
spectral slope of the late afterglow. On the other hand, when we apply the
cosmological shift to $z=1$, the $dR_C$ shift would become smaller. We
find $dR_C=-4.26$ mag for $\beta=0.5$ and $dR_C=-3.10$ mag for
$\beta=-1/3$ (also correcting for the small host extinction), resulting in $R_C=6.93$ and $R_C=7.52$,
respectively. Thus, for $\beta=-1/3$, the extrapolated $R_C$-band
magnitude at $z=1$ of the flare would be comparable to that of GRB
990123. However, we should also note that the same argument may apply to
the flare of GRB 990123, so that the flare of GRB 050904 truly is
brighter with a high probability.

\section{Is GRB 050904 different from other long GRBs?}
\label{Sec4}

In the literature on GRB 050904, consensus is not reached on whether
GRB 050904 is, apart from its redshift, a special event. Let us review
some aspects of both the prompt emission and the afterglow:

In many ways, GRB 050904 does not differ from the typical mean values
found for GRBs. In the source frame, $T_{90}=31$ seconds, with
$\gamma$-ray emission detected up to 69 seconds after the trigger,
including the flare seen prominently in the X-rays and the optical
\citep{Boer2006, Cusumano2006a, Cusumano2006b}. The intrinsic luminosities of the
X-ray \citep{Watson2006}, the late optical (as shown in this work)
and the radio afterglow \citep{Frail2006} are typical, as are the
light curve decay parameters and the jet break time in the rest frame (Z06, K06).

But some aspects of GRB 050904 do stand out. The peak energy $E_p$
lies beyond the \emph{Swift} BAT energy range. From joint fitting
of \emph{Swift} BAT and \emph{Suzaku} WAM data, \cite{Sugita2006}
report a \emph{preliminary} $E_p=366^{+568}_{-143}$ keV, translating into $E_p=2670
^{+4144}_{-1043}$ keV in the rest frame. This peak energy is the
highest found in the compilation of 56 bursts \citep{Amati2006},
only GRB 050717 \cite[observer frame $E_p=2101^{+1934}_{-830}$ keV,
time-integrated spectrum,][]{Krimm2006} very probably has a higher
peak energy (the redshift is not known but assumed to be high).
The isotropic energy is not exactly known, \cite{Cusumano2006a, Cusumano2006b}
derive $E_{\rm iso}=6.6\times10^{53}-3.2\times10^{54}$ erg. The
upper end of the range is higher than for any other burst, including
GRB 990123 \citep{Amati2006}, and even the lower value is higher
than for most GRBs with well-determined $E_{\rm iso}$. Using a jet
break time of $t_b=2.6\pm1.0$ days, \cite{Tagliaferri2005} derive
log $E_\gamma$(erg)$=51.6-52.1$. This is the second highest
beaming-corrected energy ever found \citep{FB2005}, after GRB
050820A \citep[log $E_\gamma$(erg)$=51.7-52.2$,][]{Cenko2006}.The
X-ray afterglow is the most variable ever seen by \emph{Swift}
\citep{Cusumano2006a, Cusumano2006b, Watson2006}, implying that, in the source frame,
the central engine was active for several hours after the burst
\citep{Zou2006}. We have shown in this paper that the early
optical/NIR emission is brighter than for any other GRB detected
so far. \cite{Frail2006} derive a high circum-burst density of 680
cm$^{-3}$ from broadband afterglow modeling, noting that a density
typical for GRB environments (10 cm$^{-3}$) is ruled out by the
absence of a bright early radio afterglow.

There is no aspect of GRB 050904 that exceeds the typical GRB
parameters so much that one could claim that the progenitor of GRB
050904 was markedly different from those of all other long GRBs,
e.g. a Population III star. However, it is clear that GRB 050904 is
one of the most extreme GRBs ever observed, comparable only to GRB
990123 \citep[cf.][]{Boer2006}, and exceeding even this seminal
event in some aspects.

\section{Conclusions and Outlook}
\label{Sec5}

We have shown that the early flare of GRB 050904 detected by the TAROT
telescope is the most luminous optical/NIR transient ever detected,
exceeding even the peak magnitude of the prompt optical flash of GRB
990123. While GRB 050904
is not markedly different from other GRBs in any single aspect, it is
among the most extreme GRBs ever detected.

So far, only a single GRB at the end of the reionization era has been
detected. It is thus not possible to decide whether GRB 050904 is
typical for GRBs at a very high redshift, or if it is simple
observational bias that the first GRB to be found at $z>6$ is one of
the most extreme events ever detected. More typical events would not
have sufficient follow-up to determine the necessary parameters,
especially the redshift. The \emph{Swift} satellite, with its high
sensitivity in an energy band that is typical for highly redshifted
hard bursts and its possibility of long-duration image triggers, is a
very powerful instrument for detecting these very distant
events. Probably, more very high redshift GRBs will be detected during
the lifetime of \emph{Swift}, possibly even breaking the redshift
records for quasars \citep{Fan2003} and galaxies \citep{Iye2006}.
Indeed, a GRB at a higher redshift than GRB 050904 may have already
been found, GRB 060116 \citep{Grazian2006}\footnote{However, see also
  \cite{Piranomonte2006} and \cite{Tanvir2006}.}.

While the detection of a GRB by \emph{Swift} is the first and most
important step, most parameters characterizing the event can only be
obtained by ground-based telescopes. The determination of the true
burst energetics depends on the measurement of the redshift and the
jet break time, and GRBs at $z>6$ are bright only in the NIR. While
optical spectroscopy is still feasible to $z\approx7$, any higher
redshifts will move the metal lines that are used for an exact
determination of $z$ beyond the typical limit of optical spectroscopy
(and the region beyond $0.7\;\mu$m is already strongly
affected by sky lines). NIR spectroscopy even at large facilities
needs bright sources, and thus far, NIR spectroscopy of a GRB
afterglow has never been successful in detecting any lines. The dilemma is that observing
time at large facilities is precious and observations would only be
triggered in case there is already a strong suspicion that the event
lies at a very high redshift. This implies the need for rapid
photometric redshift determinations that use the Lyman $\alpha$
trough, as it has been successfully applied to GRB 050904
\citep{Haislip2006, Tagliaferri2005, Price2006}. One telescope
that is already in use and has the capability for rapid multi-color
photometry is the MAGNUM telescope, which also observed GRB 050904
\citep{Price2006}. A key element would be the deployment of more
moderately large robotic NIR telescopes such as PAIRITEL
\citep{Bloom2005} and REM \citep{Zerbi2001}, which also has an
optical camera.

More rapid follow-up of high redshift bursts will finally allow GRBs
to fulfill their promise as the ultimate probes of the very early
universe.

\begin{acknowledgements}

We thank Jochen Greiner for useful comments and Bruce Gendre for a
very constructive referee report. D.A.K. and S.K. acknowledge
support by DFG grant Kl 766/13-2 and by the German Academic Exchange
Service (DAAD) under grant No. D/05/54048. N.M. acknowledges support
under CRUI Vigoni programme no. 506-2005.

\end{acknowledgements}



\begin{thebibliography}{}

\bibitem[Akerlof et al.(1999)]{Akerlof1999} Akerlof, C. et al. 1999, \nat, 398, 400
\bibitem[Amati(2006)]{Amati2006} Amati, L. 2006, \mnras, 372, 233
\bibitem[Bloom et al.(2005)]{Bloom2005} Bloom, J. S., Starr, D. L., Blake, C. H., Skrutskie, M. F., \& Falco, E. E. 2005, Proc. ''Astronomical Data Analysis Software \& Systems XV'' (editors C. Gabriel, C. Arviset, D. Ponz and E. Solano), submitted (astro-ph/0511842)
\bibitem[Berger et al.(2002)]{Berger2002} Berger, E. et al. 2002, \apj, 581, 981
\bibitem[Berger et al.(2006)]{Berger2006} Berger, E. et al. 2006, \apj, submitted (astro-ph/0603689)
\bibitem[B\"oer et al.(2006)]{Boer2006} B\"oer, M. 2006, \apjl, 638, L71
\bibitem[Bolzonella et al.(2000)]{hyperz} Bolzonella, M., Miralles, J.-M., \& Pell\'o, R. 2000, \aap, 363, 476
\bibitem[Cenko et al.(2006)]{Cenko2006} Cenko, S. B. et al. 2006, \apj, in press (astro-ph/0608183)
\bibitem[Chevalier \& Li(1999)]{CL1999} Chevalier, R. A., \& Li, Z.-Y. 1999, \apjl, 520, L29
\bibitem[Cusumano et al.(2006a)]{Cusumano2006a} Cusumano, G. et al. 2006, \nat, 440, 164
\bibitem[Cusumano et al.(2006b)]{Cusumano2006b} Cusumano, G. et al. 2006, \aap, in press (astro-ph/0610570)
\bibitem[Ducati et al.(2001)]{Ducati2001} Ducati, J. R., Bevilacqua, C. M., Rembold, S. B., \& Ribeiro, D. 2001, \apj, 558, 309
\bibitem[Fan et al.(2003)]{Fan2003} Fan, X. et al. 2003, \aj, 125, 1649
\bibitem[Frail et al.(2006)]{Frail2006} Frail, D. A. et al. 2006, \apjl, 646, L99
\bibitem[Friedman \& Bloom(2005)]{FB2005} Friedman, A. S., \& Bloom, J. S. 2005, \apj, 627, 1
\bibitem[Gendre et al.(2006)]{Gendre2006} Gendre, B. et al. 2006, \aap, in press (astro-ph/0603431)
\bibitem[Grazian et al.(2006)]{Grazian2006} Grazian, A. et al. 2006, GCN 4545, http://gcn.gsfc.nasa.gov/gcn/gcn3/4545.gcn3
\bibitem[Haislip et al.(2006)]{Haislip2006} Haislip, J. B. et al. 2006, \nat, 440, 181
\bibitem[Ibata et al.(1999)]{Ibata1999} Ibata, R. A., Lewis, G. F., Irwin, M. J., Leh\'ar, J. \& Totten, E. J. 1999, \aj, 118, 1922
\bibitem[Inoue et al.(2004)]{Inoue2004} Inoue, A. K., Yamazaki, R., \& Nakamura, T. 2004, \apj, 601, 644
\bibitem[Iye et al.(2006)]{Iye2006} Iye, M. et al. 2006, \nat, 443, 186
\bibitem[Kann et al.(2006)]{PaperIII} Kann, D. A., Klose, S., \& Zeh, A. 2006, \apj, 641, 993 (K06)
\bibitem[Kawai et al.(2006)]{Kawai2006} Kawai, N. et al. 2006, \nat, 440, 189
\bibitem[Krimm et al.(2006)]{Krimm2006} Krimm, H. A. et al. 2006, \apj, 648, 1117
\bibitem[Lamb \& Reichart(2000)]{Lamb2000} Lamb, D. Q., \& Reichart, D. E. 2000, \apj, 536, 1
\bibitem[Liang \& Zhang(2006)]{LZ2006} Liang, E., \& Zhang, B. 2005, \apjl, 638, L67
\bibitem[Maiolino et al.(2004)]{Maiolino2004} Maiolino, R. et al. 2004, \nat, 431, 533
\bibitem[Nardini et al.(2006)]{Nardini2006} Nardini, M. et al. 2006, \aap, 451, 821
\bibitem[Oke \& Gunn(1983)]{ABMag} Oke, J. B., \& Gunn, J. E. 1983, \apj, 266, 713
\bibitem[Pei(1992)]{Pei1992}Pei, Y. C. 1992, \apj, 395, 130
\bibitem[Piranomonte et al.(2006)]{Piranomonte2006} Piranomonte, S. et al. 2006, GCN 4583, http://gcn.gsfc.nasa.gov/gcn/gcn3/4583.gcn3
\bibitem[Price et al.(2002)]{Price2002} Price, P. A. et al. 2002, \apjl, 572, L51
\bibitem[Price et al.(2006)]{Price2006} Price, P. A. et al. 2006, \apj, 645, 851
\bibitem[Sandage(1988)]{Sandage1988} Sandage, A. 1988, \araa, 26, 561
\bibitem[Sari et al.(1998)]{SPN1998}Sari, R., Piran, T., \& Narayan, R. 1998, \apjl, 497, L17
\bibitem[Schlegel et al.(1998)]{SFD1998} Schlegel, D. J., Finkbeiner, D. P., \& Davis, M. 1998, \apj, 500, 525
\bibitem[Spergel et al.(2003)]{Spergel2003} Spergel, D. N. et al. 2005, \apjs, 148, 175
\bibitem[Stratta et al.(2005)]{Stratta2005} Stratta, G. et al. 2005, \aap, 441, 83
\bibitem[Sugita et al.(2006)]{Sugita2006} Sugita, S. et al. 2006, poster shown at the meeting of the Royal Astronomical Society, London, September 2006
\bibitem[Tagliaferri et al.(2005)]{Tagliaferri2005} Tagliaferri, G. et al. 2005, \aap, 443, L1
\bibitem[Tanvir et al.(2006)]{Tanvir2006} Tanvir, N., Levan, A. J., Priddey, R. S., Fruchter, A. S., \& Hjorth, J. 2006, GCN 4602, http://gcn.gsfc.nasa.gov/gcn/gcn3/4602.gcn3
\bibitem[Totani et al.(2006)]{Totani2006} Totani, T. et al. 2006, \pasj, 58, 485
\bibitem[Vestrand et al.(2005)]{Vestrand2005} Vestrand, W. T. et al. 2005, \nat, 435, 178
\bibitem[Vestrand et al.(2006)]{Vestrand2006} Vestrand, W. T. et al. 2006, \nat, 442, 172
\bibitem[Watson et al.(2006)]{Watson2006} Watson, D. et al. 2006, \apjl, 637, L69
\bibitem[Wei et al.(2006)]{Wei2006} Wei, D. M., Yan, T., \& Fan, Y. Z. 2006, \apjl, 636, L69
\bibitem[Wu et al.(2005)]{Wu2005}Wu, X. F., Dai, Z. G., Huang, Y. F., Lu, T. 2005, \apj, 619, 968
\bibitem[Zeh et al.(2006)]{PaperII} Zeh, A., Klose, S., \& Kann, D. A. 2006, \apj, 637, 889 (Z06)
\bibitem[Zerbi et al.(2001)]{Zerbi2001} Zerbi, R. M. et al. 2001, Astron. Nachr., 322, 275
\bibitem[Zhang \& M\'esz\'aros(2004)]{ZhangMeszaros2004} Zhang, B., \& M\'esz\'aros, P. 2004, Int.J.Mod.Phys. A19 (2004), 2385
\bibitem[Zou et al.(2006)]{Zou2006} Zou, Y. C., Xu, D., \& Dai, Z. G. 2006, \apjl, 646, 1098
\end{thebibliography}
\end{document}